\documentclass[letterpaper, preprint]{aastex}
\usepackage{amsmath, amsfonts, bbm, bm, calc}
\usepackage{graphicx}
\usepackage[normalem]{ulem}

\usepackage{algorithm, algorithmicx, algpseudocode}

\newcounter{address}
\usepackage{color}
\newcommand{\latin}[1]{\emph{#1}}
\newcommand{\etal}{\latin{et\,al.}}

\newcommand{\unit}[1]{\mathrm{#1}}
\newcommand{\bth} {\boldsymbol \theta}

\newcommand{\md}{\mathrm{d}}

\newcommand{\ee}{\mathrm{E}}

\definecolor{red}{rgb}{1,0,0}
\definecolor{blue}{rgb}{0,0,1}
\definecolor{darkgreen}{rgb}{0,0.5,0}

\usepackage{hyperref}
\usepackage{algpseudocode}

\begin{document}
\title{
  The Probabilities of Orbital-Companion Models for Stellar Radial Velocity Data
}

\author{
  Fengji~Hou\altaffilmark{\ref{CCPP}},
  Jonathan~Goodman\altaffilmark{\ref{Courant}},
  David~W.~Hogg\altaffilmark{\ref{CCPP},\ref{MPIA}}
}

\setcounter{address}{1}
\altaffiltext{\theaddress}{\stepcounter{address}\label{CCPP} Center
  for Cosmology and Particle Physics, Department of Physics, New York
  University, 4 Washington Place, New York, NY 10003}
\altaffiltext{\theaddress}{\stepcounter{address}\label{Courant}
  Courant Institute of Mathematical Sciences, New York University, 251
  Mercer Street, New York, NY 10012}
\altaffiltext{\theaddress}{\stepcounter{address}\label{MPIA}
  Max-Planck-Institut f\"ur Astronomie, K\"onigstuhl 17, D-69117
  Heidelberg, Germany}

\begin{abstract}
The fully marginalized likelihood, or Bayesian evidence, is of great importance in probabilistic data analysis, because it is involved in calculating the posterior probability of a model or re-weighting a mixture of models conditioned on data. It is, however, extremely challenging to compute. This paper presents a geometric-path Monte Carlo method, inspired by multi-canonical Monte Carlo to evaluate the fully marginalized likelihood. We show that the algorithm is very fast and easy to implement and produces a justified uncertainty estimate on the fully marginalized likelihood. The algorithm performs efficiently on a trial problem and multi-companion model fitting for radial velocity data. For the trial problem, the algorithm returns the correct fully marginalized likelihood, and the estimated uncertainty is also consistent with the standard deviation of results from multiple runs. We apply the algorithm to the problem of fitting radial velocity data from HIP 88048 ($\nu$ Oph) and Gliese 581. We evaluate the fully marginalized likelihood of 1, 2, 3, and 4-companion models given  data from HIP 88048 and various choices of prior distributions. We consider prior distributions with three different minimum radial velocity amplitude $K_{\mathrm{min}}$. Under all three priors, the 2-companion model has the largest marginalized likelihood, but the detailed values depend strongly on $K_{\mathrm{min}}$. We also evaluate the fully marginalized likelihood of 3, 4, 5, and 6-planet model given data from Gliese 581 and find that the fully marginalized likelihood of the 5-planet model is too close to that of the 6-planet model for us to confidently decide between them.

\end{abstract}


\section{Introduction}

Many inference problems take the form of model selection.
The example we focus on here is using radial velocity (RV) data to determine how many planets
orbit a given star.
A $k$-planet fit is a representation
\begin{equation}
v(t) = \sum_{i=1}^k v_i(t) + v_0 + \mbox{ noise} \; .
\label{eq:kp}  \end{equation}
A more precise version of this formula is (\ref{eq:vt}) below.
We wish to infer the ``parameter'' $k$ from RV data.
The Bayesian approach does not select a specific $k$, but gives posterior probabilities.

Bayesian model selection hinges on  the {\em fully marginalized likelihood} integral (FML),
also called the {\em Bayesian evidence} integral.
In the abstract formulation, there is data, $\cal D$, and a family of models $\mathrm{m}_k$.
The prior probability of model $\mathrm{m}_k$ is $P(\mathrm{m}_k)$.
The probability of observing data $\cal D$ is the FML, denoted $P({\cal D}\mid\mathrm{m}_k)$.
The posterior probability of model $k$ is 
\begin{equation}
P(\mathrm{m}_k\mid{\cal D}) =
\frac{P({\cal D}\mid\mathrm{m}_k)\,P(\mathrm{m}_k)}
        {\sum_j{P({\cal D}\mid\mathrm{m}_j)\,P(\mathrm{m}_j)}}\; .
\label{eq:Bayesian-decision-theory}
\end{equation}
In general, model $\mathrm{m}_k$ has parameters $\bth_k = (\theta_1,\ldots,\theta_{d_k})$.
In our planet-selection problem, there are two overall parameters per data source and 5 parameters describing the orbit 
of each planet, so $d_k = 2s+5k$, where $s$ is the number of data sources.
The prior probability density for $\bth_k$ is $\pi_k(\bth_k)$.
The probability density for the data $\cal D$ in model $k$ with parameters $\bth_k$ is 
the {\em likelihood function}, $L_k({\cal D}\mid \bth_k)$.
The overall probability of $\mathrm{m}_k$ is the integral over all possible parameter values
\begin{equation}
P({\cal D}| {\mbox m}_k) = \int L_k({\cal D}|\bth_k)\, \pi_k(\bth_k) \,\md\bth_k \; .
\label{eq:FML}  \end{equation}
This paper suggests a way to compute this challenging integral.
Other approaches include reversible jump MCMC \citep{richardson97a}, 
parallel tempering \citep{ford07a}, nested sampling \citep{skilling06a, feroz08a}, diffusive 
nested sampling \citep{brewer11a}, and population Monte Carlo \citep{kilbinger10a}.
Our method has computational advantages over these in the case in which the data are good enough such that the
posterior parameter distribution, 
$P(\bth_k | {\cal D}, \textrm{m}_k)\propto L_k({\cal D}|\bth_k)\, \pi_k(\bth_k)$ is much more localized 
than the prior distribution, $\pi_k$.

The generic evidence integral FML problem is to evaluate integrals of the form
\begin{equation}
Z = \int\! L(\bth)\,\pi(\bth)\,\md\bth\; .
\label{eq:evidence-integral}
\end{equation}
We assume it is possible to evaluate $\pi$ and $L$.
We find an approximation $g(\bth)$ of the integrand and consider a {\em geometric path}, 
$g(\bth)^{1-\beta}(L(\bth)\,\pi(\bth))^\beta$.
The corresponding integrals are
\begin{equation}
Z_{\beta} = \int g(\bth)^{1-\beta}(L(\bth)\,\pi(\bth))^\beta\, d\bth \; .
\label{eq:Zb}  \end{equation}
The starting integral $Z_0$ can be known in closed form or easy to evaluate.
The desired integral is $Z_1$.
Our method is motivated by the multi-canonical Monte Carlo approach used in statistical physics to 
evaluate the partition function as a function of temperature,  \citep{berg99a}.
We are also inspired by \cite{gelman98a} to use geometric path, but we differ from them that we do not sample paths.
For our problems, we use a Gaussian approximation to the posterior as $g(\bth)$.
In this case, $Z_0$ is what the {\em Bayesian information criterion} (BIC)  uses as an estimate of 
the desired $Z_1$ as a model selection criterion \citep{bishop07a}.

Variance estimation and error bars are an essential part of any Monte Carlo computation.
Our algorithm chooses steps $\Delta \beta$ using explicit MCMC variance estimates that use estimated
auto-correlation times.
This leads to a robust algorithm and estimates of $Z$ with explicit uncertainty estimates or error bars.
The estimated error bars agree with error bars from experiments with multiple independent evaluations. 

The basic geometric path idea goes back at least to the multi-histogram thermodynamic 
integration method of \cite{ferrenberg89a}, which was our motivation.
Adaptions for computing the FML were developed by several authors, see \cite{xie11a}, \cite{fan11a}, and
references there.
We differ from them in certain technical but important ways. 
We choose the steps $\Delta\beta$ (see below for notation and definitions) adaptively during the 
computation using on-the-fly estimates of the variance as a function of $\Delta \beta$.
This leads to efficient and robust evaluations with specified error bars.
Also, we choose the starting distribution $p_0$ as multivariate Gaussian whose covariance 
matches the estimated covariance matrix of the posterior.
It seems to make a large difference for our applications, in which the components of $\theta$ are highly
correlated in the posterior, to get the covariance structure of the posterior right from the beginning.

We apply our algorithm to evaluating the FML of multi-companion models fitting for RV data from HIP 88048 ($\nu$ Oph) taken as part of the Lick K-Giant Search \citep{frink02a, mitchell03a, hekker06a, hekker08a, quirrenbach11a}. 
We choose HIP 88048 to study because it has two confirmed brown-dwarf companions of approximately 530-d period and 3210-d period \citep{quirrenbach11a}, but may hide additional companions. 
Also the noise level of HIP 88048 is low, so over-fitting should be more obvious when excessive amount of companions are added to the model. 
We evaluate the FML of 1, 2, 3 and 4-companion models based on the data. 
The result shows that 2 companion model has the largest FML among them given various prior distributions.
We also apply the algorithm to evaluate the FML of multi-planet models for Gliese 581.
We use the combination of both HARPS data \citep{mayor09a} and HIRES data \citep{vogt10a}.
We evaluate the FML of 3, 4, 5, and 6-planet models.
We find that the 5 and 6-planet models have the largest FMLs among the four models.
But the FML difference between 5 and 6-planet model is much smaller than reported in \cite{gregory11a}.

We point out two philosophical issues: the criteria for model selection, and the role of the prior.
There are (at least) two reasons to do model selection.
One is to estimate the number of planets or companions of a given star above a threshold size.
Another is to accurately predict future measurements.
These can lead to different model selection criteria and different results.
A model with fewer than the correct number of planets has fewer parameters and therefore may suffer less
from over-fitting.
Cross validation is a model selection criterion based on measuring predictive power,
while Bayesian FML criteria are based on Bayesian estimates of the number of planets.

\begin{figure}
 \includegraphics[width=0.99\linewidth]{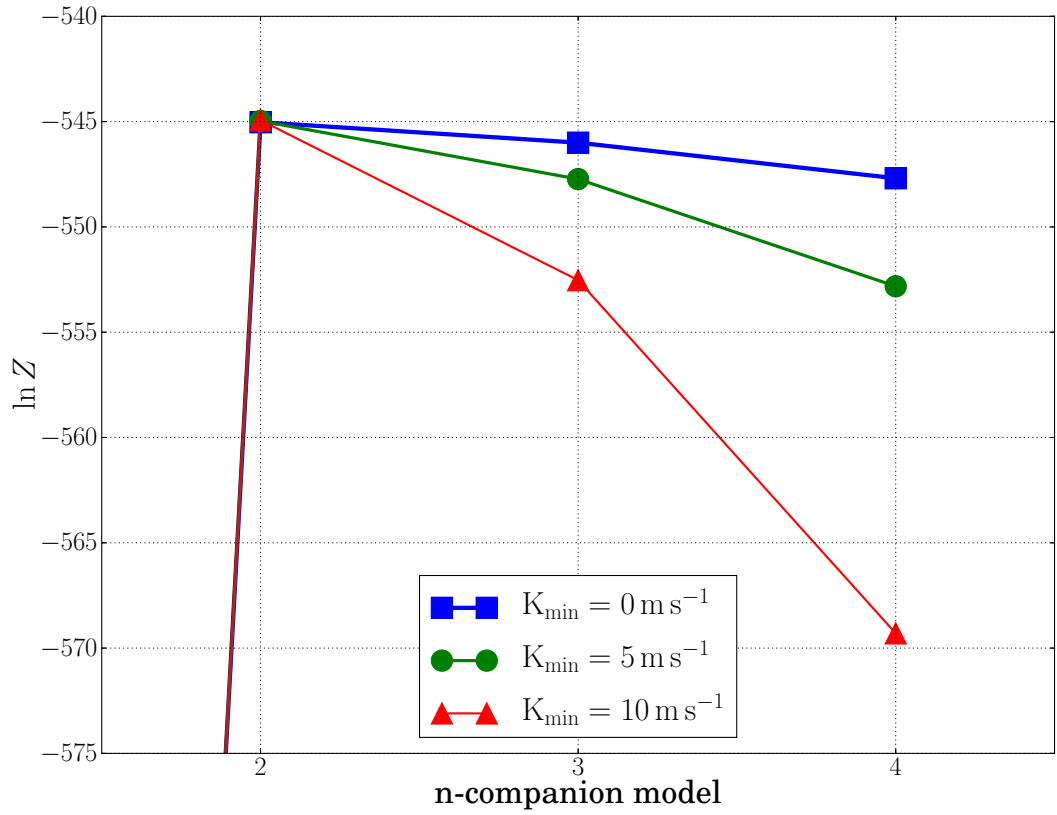}
 \caption{The logarithm of the marginalized likelihoods of 2, 3 and 4-companion models given RV data from HIP 88048 and three different $K_{\mathrm{min}}$.}
 \label{fig:282lnZK}
\end{figure}

The need to specify somewhat subjective priors is a weakness of Bayesian estimation.
Priors contain normalization factors that depend on things such as the range of allowed amplitudes.
In cases where the prior is uncertain, Bayesians hope that the posterior is robust with respect to 
details of the prior.
The present model selection results are less robust with respect to the prior, see Fig.~(\ref{fig:282lnZK}).
As a simple illustration, suppose each planet has a parameter whose prior is uniformly distributed 
in a range $[0,R]$.  
The prior density for each planet has a normalization factor $R^{-1}$.
For $k$ planets with independent priors, that gives an overall factor $R^{-k}$.
Suppose further that the data lead to a likelihood that is almost zero when this very parameter is 
in the range $[R/2,R]$.
Then the posterior is almost the same if we assume a smaller range $[0,R/2]$, so the posterior is 
robust with respect to the prior.
But the FML integrals for different $k$ values are changed by factors $2^k$, which are on the 
order of the differences reported in Fig.~(\ref{fig:282lnZK}).
This is true even if all of the models essentially exclude the range $[R/2,R]$.
See \cite{kilbinger10a} for a different thoughtful discussion of priors in model selection.

There are many aspects of the multi-planet priors that need to be examined.
In the present study of HIP 88048, we focus only on the possibility of small planets.
All stars may have some satellites.  
It may be a more interesting question to ask how many planets there are above a given observable size.
The example of Fig.~(\ref{fig:282lnZK}) shows that the data slightly favor two planets to three 
with no lower constraint on the size.
But with a constraint of at least $10\,\mathrm{m\,s^{-1}}$, the relative weights differ by a 
factor of $e^7 \sim 10^3$.

Section~\ref{sec:m-cmc} explains the geometric-path Monte Carlo algorithm that we use. 
Section~\ref{sec:rosenbrock} describes a simple model computation that allows us to validate the 
algorithm, both the computed answer and the estimated variance.
Section~\ref{sec:tszyj} presents FML computations for the interesting star HIP 88048.
We are able to conclude that it probably has two significant planets even though there are suggestions 
of a third.
Any planets beyond the two confirmed ones are too small to be identified using the RV data we have.
Finally, In Section~\ref{sec:discussion}, we discuss our findings and future projects. 

\section{Geometric-Path Monte Carlo}
\label{sec:m-cmc}
\begin{figure}
 \includegraphics[trim = 4cm 0 0 0, width=1.1\linewidth]{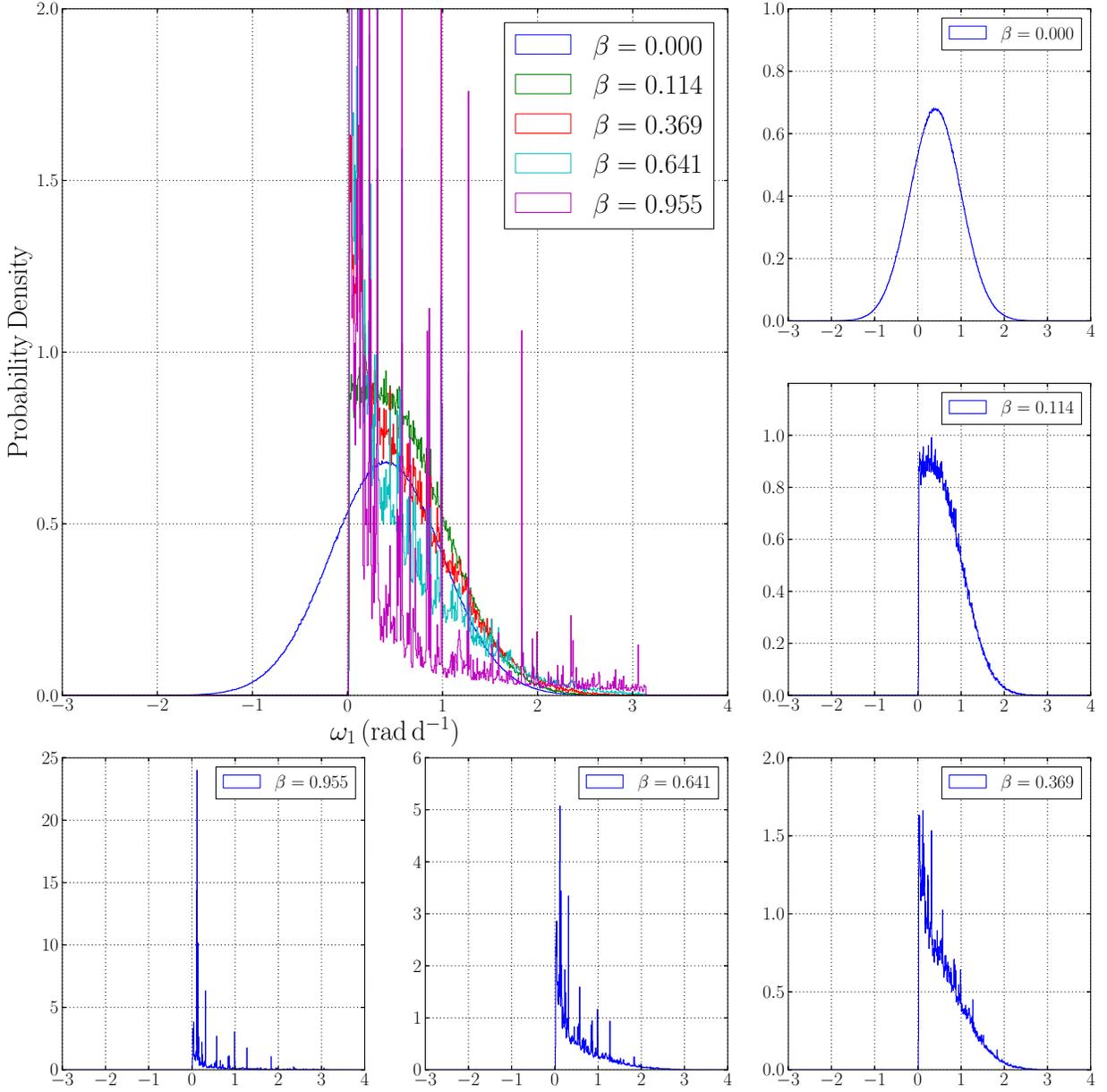}
 \caption{Histograms for $p_{\beta}(\omega_1)$ for several values of $\beta$.
The histograms shown in this figure are those for the $ \omega_1$ in 3-companion model for HIP 88048. 
On the right column and bottom row are individual histograms for each $\beta$.
The large picture on the left top is the combination of all these histograms.
The vertical limits are set between 0 and 2 to make the contrast more visible.
The $\beta = 0$ histogram is Gaussian. 
The $\beta = 0.955$ histogram is very similar to the posterior of $\omega_1$ shown in Fig.~(\ref{fig:282-3-hist}).
All the peaks in these histograms are fully resolved. }
 \label{fig:282m3gmc}
\end{figure}

To implement the geometric-path Monte Carlo algorithm, we first find a distribution $g(\bth)$, which is an approximation to the posterior distribution and the normalization of which is known exactly.
We then define a mixture of $g(\bth)$ and the posterior distribution according to the geometric path,
\begin{equation}
p_\beta(\bth) \equiv \frac{1}{Z_\beta}\, (L(\bth)\,\pi(\bth))^\beta\, g(\bth)^{(1-\beta)}\,,
\label{eq:p-beta}
\end{equation}
where $Z_\beta$ is a normalization factor, and $\beta$ labels the geometric path and ranges between 0 and 1. 
Eqn.~(\ref{eq:p-beta}) is similar to the geometric path used in bridge sampling or path sampling \citep{gelman98a}.
It is easy to see that
\[
p_0(\bth) = g(\bth)\, ,
\]
and
\[
p_1(\bth) = \frac{1}{Z_1}\,L(\bth)\,\pi(\bth)\, ,
\]
where $Z_1$ is $Z$, the FML. Note that $Z_0=1$ is known exactly.
Fig.~(\ref{fig:282m3gmc}) illustrates the change of $p_\beta$ with $\beta$. 
We estimate $Z_{\beta_k}$ for an increasing sequence $\beta_{k+1} = \beta_k +\Delta \beta_k$, starting
with $\beta_1=0$, and continuing to $\beta_L = 1$.
If $Z_k$ is known, then $Z_{\beta_k+\Delta \beta}$ is found from $Z_{\beta_k}$ using
\begin{align}
Z_{\beta_k+\Delta \beta}
&= \int \left(L(\bth)\,\pi(\bth)\right)^{\beta_k+\Delta\beta}\,g(\bth)^{(1-\beta_k-\Delta\beta)}\,\md\bth \nonumber \\
&= Z_{\beta_k} \int \left( \frac{ L(\bth) \, \pi(\bth) }{ g(\bth) } \right)^{\Delta\beta} \, p_{\beta_k}(\bth) \, \md \bth  \label{eq:epk} \\
&= Z_{\beta_k} \, W(\beta_k, \Delta\beta)\, .
\label{eq:Z-beta_2}
\end{align}
The integral in (\ref{eq:epk}) expresses $W$ as an expectation under $p_{\beta_k}$.
We write this as
\begin{equation}
W(\beta_k, \Delta\beta) = \ee_{\beta_k} \left[ Y(\bth)^{\Delta\beta}  \right]\, ,
\end{equation}
where
\begin{equation}
Y(\bth) \equiv  \frac{ L(\bth)\,\pi(\bth) }{ g(\bth) } \, .
\label{eq:Y-sample}
\end{equation}
The representations (\ref{eq:Z-beta_2}) may be combined to yield the product formula
\begin{equation}
Z = Z_{\beta_L} = 
W(\beta_1,\Delta \beta_1) W(\beta_2,\Delta \beta_2) \cdots W(\beta_{L-1},\Delta \beta_{L-1}) \; .
\label{eq:Zpr}  \end{equation}
We estimate the factors $W(\beta_k,\Delta \beta_k)$ using MCMC sampling of $p_{\beta_k}$ then multiply
these estimates to estimate $Z$.

In our method, we choose a Gaussian distribution as $g(\bth)$, and it can be expressed as
\begin{equation}
g(\bth)\equiv\frac{\sqrt{\det{H}}}{(2\,\pi)^{\md/2}}\exp{\left(-\frac{(\bth-\textbf{m})^T H (\bth-\textbf{m})}{2}\right)}\,.
\label{eq:gaussian}
\end{equation}
where $\md$ is the dimension, $\textbf{m}$ the mean vector and $H$ the inverse of the covariance of the Gaussian.
There are various ways to find a suitable $\textbf{m}$ and $H$. 
One could take $\textbf{m}$ to be the global maximum of $L(\bth) \, \pi(\bth)$ and $H$ to be the Hessian 
matrix of $\log(L(\bth) \, \pi(\bth))$ at $\bth=\textbf{m}$. 
This corresponds to using the Laplace integration approximation to the evidence integral
(\ref{eq:evidence-integral}), see e.g., \citep{murray84a}, which is the basis of the {\em BIC}, or 
{\em Bayesian Information Criterion}, see e.g., \citep{bishop07a}.
Our approach uses less computational infrastructure.
We use an MCMC sampler to sample the posterior, then take $\textbf{m}$ to be the
empirical mean and $H$ the inverse of the empirical covariance matrix of these samples. 

The MCMC sampler that we use to sample the posterior and $p_k$ is the affine invariant ensemble sampler by stretch move from the emcee package \citep{foreman-mackey13a}.
This algorithm has the advantage of being able to sample highly anisotropic distributions without problem-dependent tuning \citep{goodman10a, hou12a}.

We use the MCMC estimator of $W(\beta_k, \Delta\beta)$, which is 
\begin{equation}
\widehat{W}(\beta_k, \Delta\beta) = \frac{1}{N} \sum_{i=1}^N { Y(\bth_i)^{\Delta\beta} }\; ,
\label{eq:W-est}
\end{equation}
where $\bth_i$ are samples from $p_{\beta_k}$.
Define $V(\beta_k, \Delta\beta)$ as the variance of $\widehat{W}(\beta_k, \Delta\beta)$. 
We use the variance estimator described in \citep{sokal}, which is
\begin{equation}
\widehat{V}(\beta_k, \Delta\beta) = \frac{\widehat{\tau}(\beta_k)}{N} 
\sum_{i=1}^N \left( { Y(\bth_i)^{\Delta\beta}} - \widehat{W}(\beta_k, \Delta\beta) \right)^2\, ,
\label{eq:V-est}
\end{equation}
where $\widehat{\tau}(\beta_k)$ is an estimate of the auto-correlation time of the chain of 
$Y(\bth_i)$, estimated using the self-consistent window method.
In principle (see \citealt{sokal}), we should use the autocorrelation time of $Y(\bth_i)^{\Delta\beta}$.
But this is more expensive to compute.
We hope $\tau$ is not a strong function of the exponent $\Delta \beta$.
The estimation error, $\widehat{W} - W$, is likely to be of the order of its standard deviation, $\sqrt{V}$.
The {\em relative error} is the estimation error normalized by the quantity being estimated.
A standard estimate of the relative error is
\begin{equation}
R = \frac{\sqrt{\widehat{V}(\beta_k,\Delta\beta)}}{\widehat{W}(\beta_k, \Delta\beta)} \;.
\label{eq:variance-control}
\end{equation} 

We choose $\Delta \beta_k$ to be the largest $\Delta \beta$ with $R \leq C$,
where $C$ is a pre-set value. 
It is possible to do this because
\begin{equation}
\widehat{V}(\beta_k, \Delta\beta) \to 0 \;\;\;\mathrm{as}\;\Delta\beta \to 0\,  ,
\label{eq:limit-V}
\end{equation}
while
\begin{equation}
\widehat{W}(\beta_k, \Delta\beta) \to 1 \;\;\;\mathrm{as}\;\Delta\beta \to 0\,  ,
\label{eq:limit-W}
\end{equation}
as long as there are no 0's in $Y(\bth_i)$. 
The situation that the chain of $Y(\bth_i)$ contains 0 usually happens when $\beta_1=0$ and one is sampling $g(\bth)$. 
Because there may be samples from $g(\bth)$ which have zero prior or likelihood, the numerator in Eqn.~(\ref{eq:Y-sample}) is then zero. 
In such cases, there is a non-zero lower limit of $R$, which we denote as $C_0$. 
If $C_0>C$, one can take a small enough $\Delta\beta$ as $\Delta\beta_k$, so that $R$ achieves the lower limit $C_0$. 
Both $C$ and $N$ are tuning parameters of this algorithm. 
How to set them depends on specific problems. 
One may need to do a few trial runs to set $C$ and $N$ values that are sensible. 

It is convenient to describe the above MCMC estimation errors in the following way.
Given enough burn-in, the MCMC estimates are nearly unbiased.
For large $N$, they are approximately normal.
We choose $\Delta \beta_k$ so that the error standard deviation is roughly $CW_k$.
Therefore, we may use the approximate error expression $\widehat{W}_k = W_k + CW_k \varepsilon_k$, where
the $\varepsilon_k$ are ``standard'' independent Gaussians with mean zero and variance one.

We use the natural estimator of the product (\ref{eq:Zpr}), which is 
\begin{equation}
\widehat{Z} = \prod_{k=1}^{M-1} \widehat{W}(\beta_k,\Delta \beta_k)\,  . 
\end{equation}
With the above error description, this may be approximated as
\begin{align*}
\prod_{k=1}^{M-1} \widehat{W}(\beta_k,\Delta \beta_k) 
&= \prod_{k=1}^{M-1} W(\beta_k,\Delta \beta_k)\prod_{k=1}^{M-1} \left(1+C\varepsilon_k\right) \\
&= Z \left( 1 + C \sum_{k=1}^{M-1} \varepsilon_k \right)\; .
\end{align*}
This leads directly to an estimate of the standard error of $\widehat{Z}$, which is 
\begin{equation}
\sigma_{\widehat{Z}} = Z\,\sqrt{M-1}\,C\, .
\label{eq:z-var-est}
\end{equation}
At the end of a run we have a good simple error bar for the computed evidence integral, because
everything on the right side of (\ref{eq:z-var-est}) is known or estimated.

\section{Trial Problem with Rosenbrock Function}
\label{sec:rosenbrock}
\begin{figure}
 \includegraphics[width=0.99\linewidth]{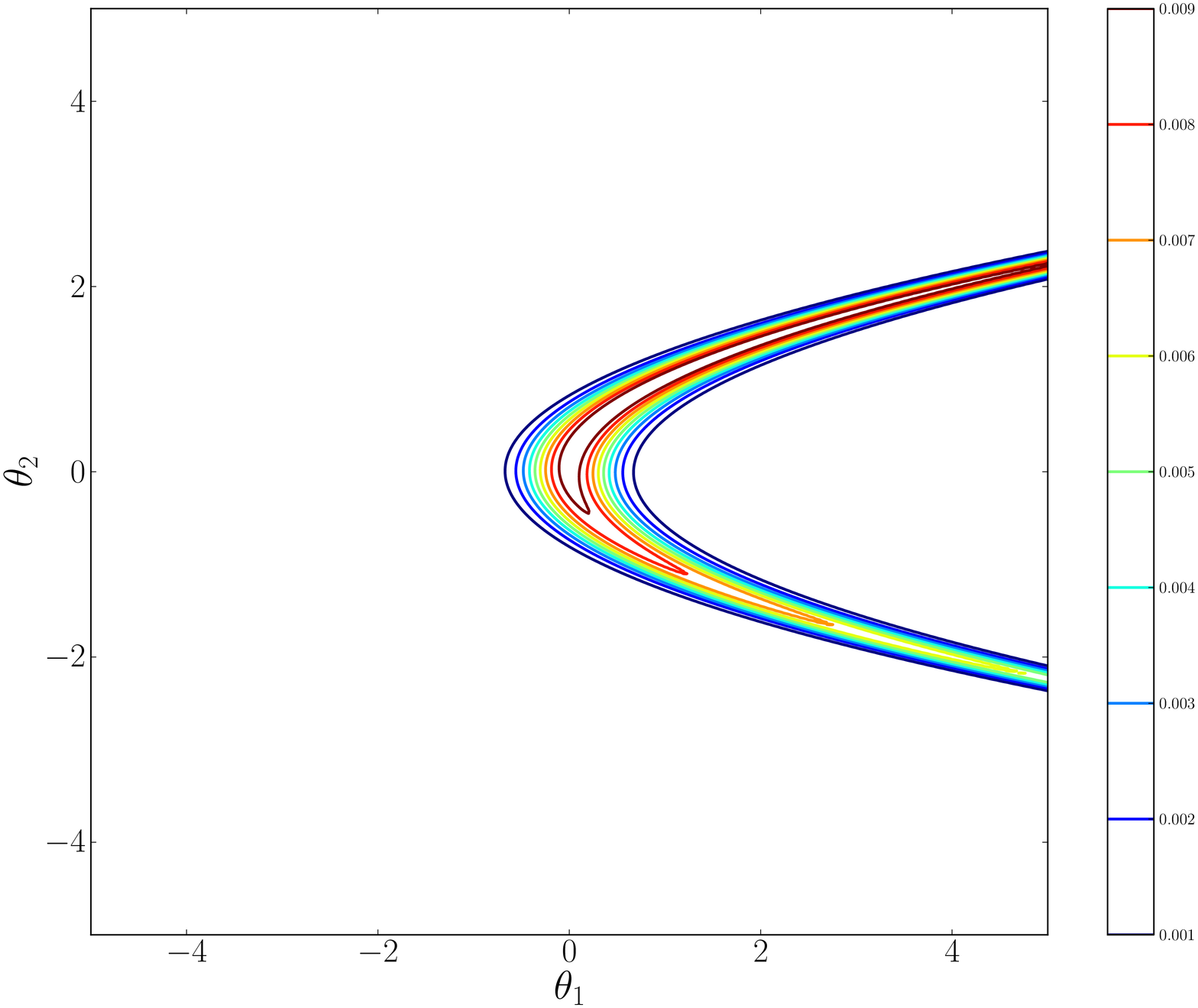}
 \caption{The contour plot of $L_R(\bth)\,\pi_R(\bth)$, defined in Eqn.~(\ref{eq:RL}) and Eqn.~(\ref{eq:Rpi}).}
 \label{fig:rosen}
\end{figure}

We apply the algorithm to an integral involving the Rosenbrock function, because the two variables in the Rosenbrock functions are highly correlated and their distributions deviate from Gaussian distribution greatly, which we hope will mimic some properties of the posterior of real problems. Specifically, we will try to evaluate the following integral
\begin{equation*}
Z_R=\int L_R(\bth)\,\pi_R(\bth)\,\md\bth\, ,
\end{equation*}
where
\begin{equation}
\pi_R(\bth) =\frac{1}{100}\, \mathbbm{1}_{[-5,\,5]}(\theta_1)\, \mathbbm{1}_{[-5,\,5]}(\theta_2)\, ,
\label{eq:Rpi}
\end{equation}
and
\begin{equation}
L_R(\bth) = \exp \left(  - \frac{100\,(\theta_2-\theta_1^2)^2 + (1-\theta_1)^2}{20} \right)\, .
\label{eq:RL}
\end{equation}
The contour plot of $L_R(\bth)\,\pi_R(\bth)$ is shown in Fig.~(\ref{fig:rosen}).
A direct quadrature with $10^4\times 10^4$ points in the square $[-5,5]\times [-5,5]$ gives 
$Z_R = 3.133\cdot 10^{-2}$.

We ran the above algorithm with $N=10^6$ samples for each $\Delta\beta_k$ and variance control parameter 
$C = 10^{-3}$. 
The mean vector and the inverse of the covariance matrix are obtained from sampling the posterior:

\begin{equation}
H = \begin{pmatrix}
               10.05 & 0.2136\\
               0.2136 & 7.798 \end{pmatrix}\;\; , \;\;\;\;
\textbf{m} = \begin{pmatrix}
               0.1572\\
               1.560 \end{pmatrix}\, .
\end{equation}

We repeat the evaluation $n = 100$ times to get multiple independent results $\widehat{Z}_1,\,\widehat{Z}_2,\,\ldots,\,\widehat{Z}_{n}$. 
The algorithm turned out to use $M=5$ values of $\beta_k$. 
The mean of the estimates was
\begin{equation}
\bar{Z} = \frac{1}{n}\sum_{i=1}^{n} \widehat{Z}_i= 0.03133\, .
\end{equation}
This agrees well with answer computed by quadrature.
The standard deviation of the $n$ evaluations is
\begin{equation}
\widehat{\sigma_{\widehat{Z}}} = \sqrt{\frac{1}{n}\sum_{i=1}^{n} (Z_i-\bar{Z})^2} = 5.6\times10^{-5}\, .
\end{equation}
This agrees reasonably well with the standard error predicted by (\ref{eq:z-var-est}), which is
$.031 \times \sqrt(4)\times 10^{-3}= 6.2\times 10^{-5}$.

\section{Multi-Companion Model Fit for HIP 88048 and Gliese 581}
\label{sec:tszyj}
HIP 88048 ($\nu$ Ophiuchi) is a K0III star, and has mass $3.04\,M_\sun$ and radius $15.1\,R_\sun$ \citep{sato12a}.
We have in total 131 radial velocity data from HIP 88048 \citep{frink02a, mitchell03a, hekker06a, hekker08a, quirrenbach11a}. 
We estimated orbital parameters for two brown-dwarf companions using the method of \citep{hou12a}.
The parameter means and standard errors are listed in Tab.~(\ref{tab:282param}).

Gliese 581 (HIP 74995) is a M3V star, and has mass $0.31\,M_\sun$ and radius $0.29\,R_\sun$ \citep{bonfils05a}.
With all the available HARPS data, \cite{mayor09a} reported a total of four planets around the star.
By combining both HAPRS and HIRES data, \cite{vogt10a} reported two additional planets.
\cite{gregory11a} did a bayesian analysis on HARPS data from Gliese 581 and reported that 5-planet model has the largest marginalized likelihood.

The 5 orbital parameters per companion are the velocity amplitude $K$, the mean angular speed $\omega$, 
the longitude of ascending node $\phi$, the eccentricity $e$, and the longitude of periastron $\varpi$.
There are two additional overall fitting parameters, the velocity offset $v_0$ and the jitter $S$.
There are $5k+2$ parameters for a $k$-companion model for HIP 88048 and $5k+4$ parameters for Gliese 581.

\subsection{Prior Distributions}
We use standard prior distributions as described in \citep{ford05a} and \citep{hou12a}.
For amplitude $K$, we have
\begin{equation}
\pi(K) = \frac{1}{\log{\frac{K_{\mathrm{max}}+K_0}{K_{\mathrm{min}}+K_0}}}\,\frac{1}{K+K_0}\, , \;\;\;\;\; K_{\mathrm{min}} < K < K_{\mathrm{max}}\, ,
\label{eq:prior-K}
\end{equation}
where we set $K_0 = 10\, \unit{m\,s^{-1}}$.
For Gliese 581, we use $K_{\mathrm{max}} = 1000\, \unit{m\,s^{-1}}$ and $K_{\mathrm{min}} = 0\, \unit{m\,s^{-1}}$.
For HIP 88048, we use $K_{\mathrm{max}} = 10000\, \unit{m\,s^{-1}}$ in order to include most substellar companions.
And we choose three different lower bounds on $K$ for HIP 88048: $K_{\mathrm{min}} = 0\, \unit{m\,s^{-1}}$, $ 5\, \unit{m\,s^{-1}}$ and $ 10\, \unit{m\,s^{-1}}$.

For angular speed $\omega$, we have
\begin{equation}
\pi(\omega) = \frac{1}{\log{\frac{\omega_{\mathrm{max}}+\omega_0}{\omega_{\mathrm{min}}+\omega_0}}}\,\frac{1}{\omega+\omega_0}\, , \;\;\;\;\; \omega_{\mathrm{min}} <\omega< \omega_{\mathrm{max}}\, ,
\end{equation}
where we set $\omega_0 = 0.01\, \unit{rad\,d^{-1}}$ and $\omega_{\mathrm{min}} = 0\, \unit{rad\,d^{-1}}$. 
For Gliese 581, we use $\omega_{\mathrm{max}} = 2\pi\, \unit{rad\,d^{-1}}$.
For HIP 88048, we use $\omega_{\mathrm{max}} = \pi\, \unit{rad\,d^{-1}}$.

For eccentricity $e$, we use a beta distribution with one of the hyper-parameters 1 and the other one 5, so that the distribution has more weight at around 0, and also goes all the way to 1. The prior for $e$ is
\begin{equation}
\pi(e) = 5\,(1-e)^4\, , \;\;\;\;\; 0 < e < 1\,.
\label{eq:prior-e}
\end{equation}

For both $\varpi$ and $\phi$, we simply use uniform distribution between $0$ and $2\pi$ as their priors. For the velocity offset $v_0$, we use uniform distribution between $-5000\, \unit{m\,s^{-1}}$ and $5000\, \unit{m\,s^{-1}}$ as its prior. 

For jitter $S$, we have
\begin{equation}
\pi(S) = \frac{1}{\log{\frac{S_{\mathrm{max}}+S_0}{S_{\mathrm{min}}+S_0}}}\,\frac{1}{S+S_0}\, , \;\;\;\;\; S_{\mathrm{min}} < S < S_{\mathrm{max}}\, ,
\end{equation}
where we choose $S_{\mathrm{min}} = 0\, \unit{m^2\,s^{-2}}$, $S_{\mathrm{max}} = 100000\, \unit{m^2\,s^{-2}}$, and $S_0 = 100\, \unit{m^2\,s^{-2}}$.

We also include a factor in the prior forbidding the orbits of the companions from crossing each other. We require that the radius of a companion in an inner orbit at apoastron is smaller than the radius of a companion in an outer orbit at periastron. So for $k$-companion model, the overall prior is
\begin{equation}
\pi(\bth) = \frac{1}{C\,k!} \mathbbm{1}_{\mbox{\scriptsize orbits not crossed}}(\bth)\, \pi(v_0)\,\pi(S)\,\prod_{i=1}^k\left[ \pi(K_i)\,\pi(\omega_i)\,\pi(\phi_i)\,\pi(e_i)\,\pi(\varpi_i)  \right] \, ,
\end{equation}
where $\mathbbm{1}$ is an indicator function, if orbits not crossed is true for $\bth$, $\mathbbm{1}$ is 1; otherwise, 0. We have a factorial $k!$ in the formula, because we require the periods of the companions to be in a monotonic order, for the sake of sampling. The coefficient $C$ is an overall normalization. 
To estimate the Bayesian evidence, which is the purpose of this paper, 
we need these normalization factors to be correct.
We have confirmed that our priors are correctly normalized by using a likelihood $L(\bth)=1$ and the geometric-path Monte Carlo described in this paper for integration. 

\subsection{Likelihood}
The likelihood function for RV data from a single source is
\begin{equation}
L(\bth)=
(2\pi)^{N_{\mathrm{data}}/2} \, \prod_{i=1}^{N_{\mathrm{data}}} \left[ (\sigma_i^2+S)^{-1/2}\, 
\exp{\left( - \frac{(v_i-v_{\mbox{\scriptsize rad}}(t_i,\bth))^2}{2(\sigma_i^2+S)}\right)} \right] \; ,
\label{eq:mpl}
\end{equation}
where $\{t_i,\,v_i,\,\sigma_i\}$ are the data, and $N_{\mathrm{data}}$ is the number of RV measurements. 
The model radial velocity is given by \citep{ohta05a},
\begin{equation}
v_{\mbox{\scriptsize rad}}(t,\bth) 
= v_0 + \sum_{i=1}^k\left[K_i\left(\sin{(f_i+\varpi_i)}+e_i\,\sin{\varpi_i}\right)\right]\, ,
\label{eq:vt}  \end{equation}
where the true anomaly $f$ is a function of $t$.
It is found by solving first for the mean anomoly $E$ in 
\begin{equation}
\omega\,t + \phi = E - e\,\cos{E}\; ,
\label{eq:mean-anomaly}
\end{equation}
then solving
\begin{equation}
\cos{f} = \frac{ \cos{E} - e}{1 - e\,\cos{E}}\;.
\label{eq:true-anomaly}
\end{equation}
We have omitted the companion indexes in Eqn.~(\ref{eq:true-anomaly}) and Eqn.~(\ref{eq:mean-anomaly}). If there are multiple data sources, there will be corresponding number of $S$ and $v_0$.

\subsection{Marginalized Likelihood}
\subsubsection{HIP 88048}
\begin{figure}
 \includegraphics[width=0.99\linewidth]{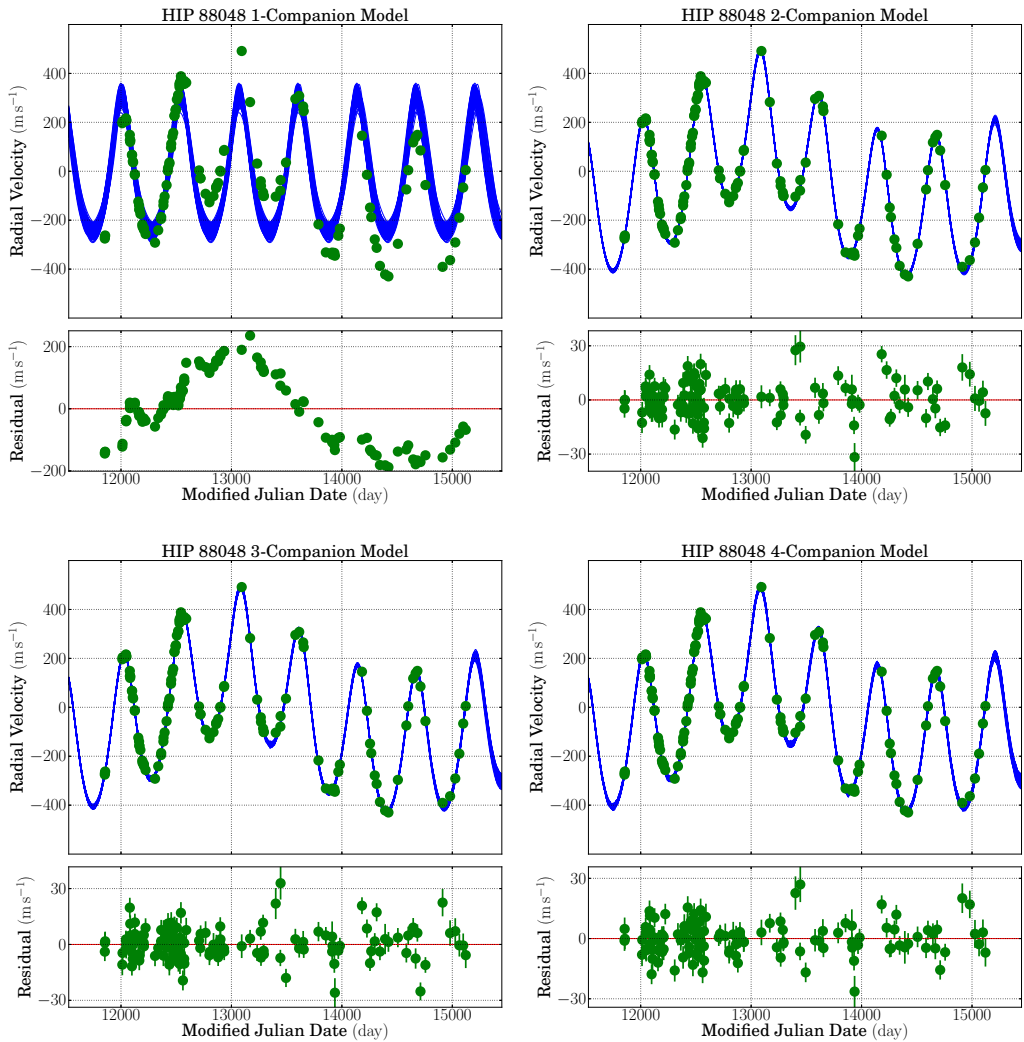}
 \caption{Fits of various models for RV data from HIP 88048 given $K_{\mathrm{min}} =0\,\unit{m\,s^{-1}}$. The upper left figure shows the fit and residual of 1 companion model. The upper right one shows those of 2-companion model. The lower left is for 3 companion model. And the lower right is for 4 companion model. 100 fits drawn from the posterior are plotted for all four models. The residual plots are of the optimum fit for all four models.}
 \label{fig:282-all-fits}
\end{figure}

\begin{figure}
 \centering
 \includegraphics[width=0.9\linewidth]{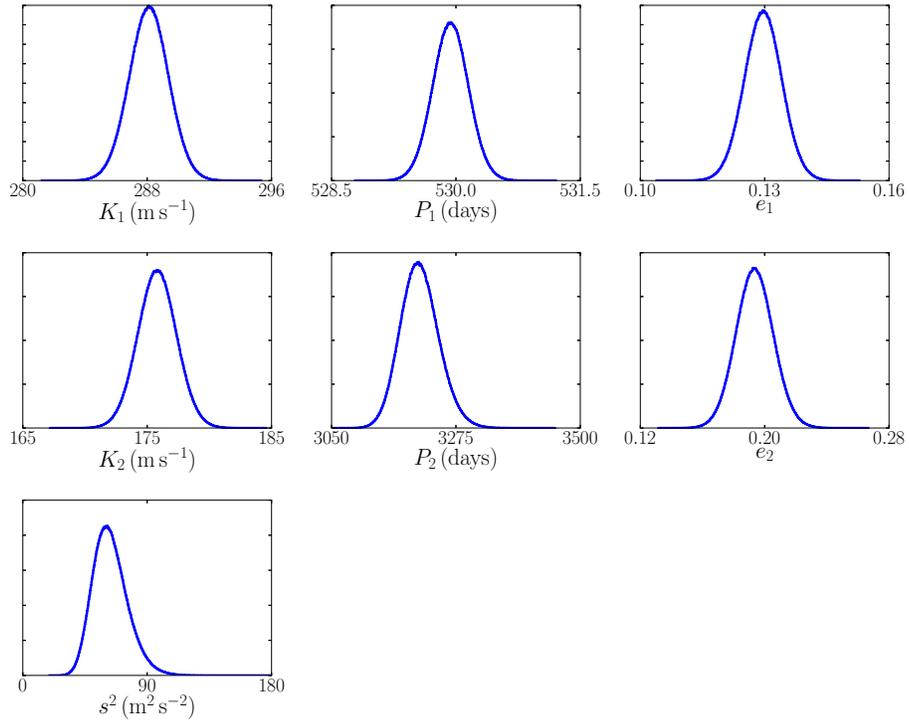}
 \caption{The posterior histograms of amplitudes, periods, eccentricities and jitter in the 2-companion model for RV data from HIP 88048 and $K_{\mathrm{min}} =0\,\unit{m\,s^{-1}}$. All the parameters are well constrained by the posterior distribution.}
 \label{fig:282-2-hist}
\end{figure}

\begin{figure}
 \centering
 \includegraphics[width=0.9\linewidth]{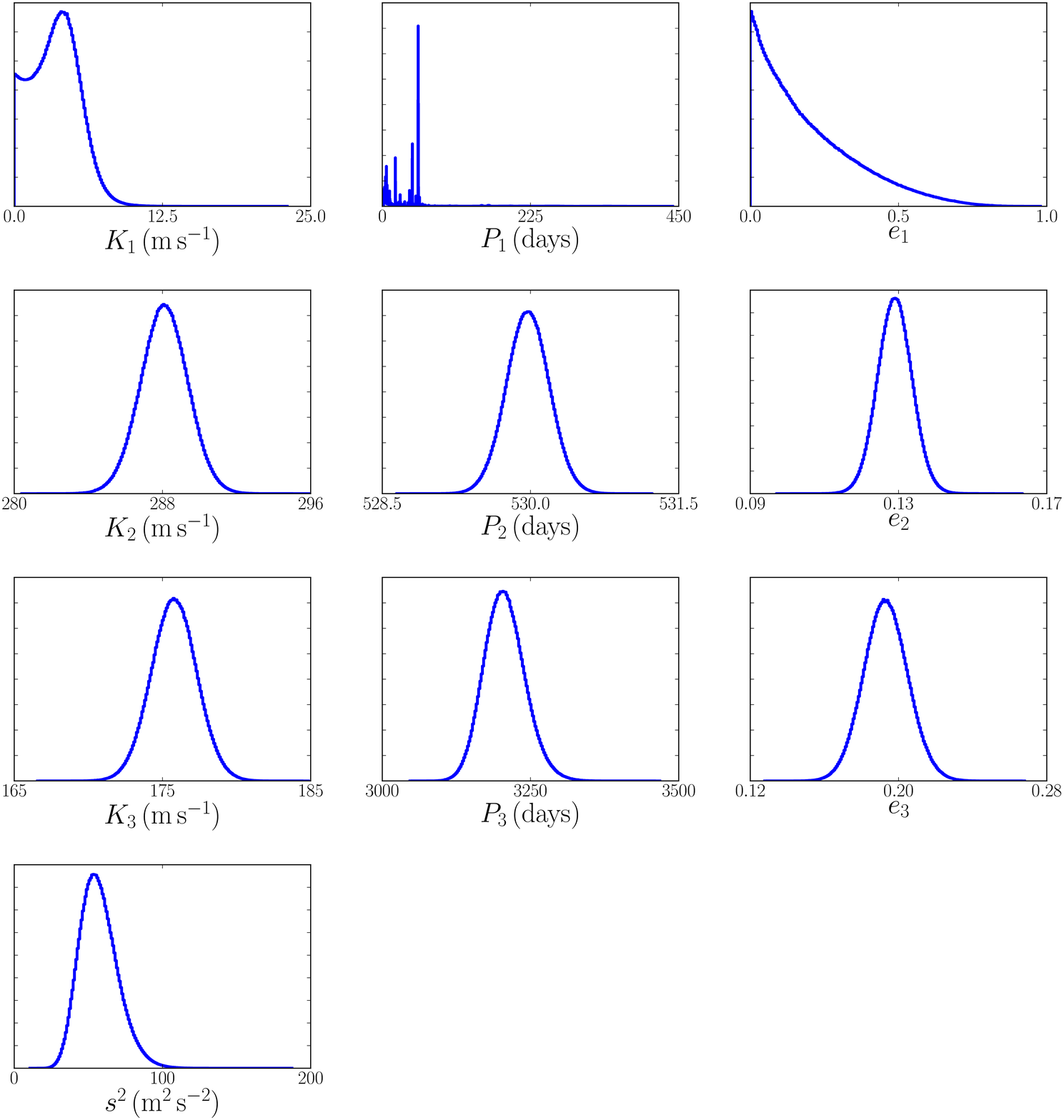}
 \caption{The posterior histograms of amplitudes, periods, eccentricities and jitter in the 3-companion model for RV data from HIP 88048 and $K_{\mathrm{min}} =0\,\unit{m\,s^{-1}}$. The histogram of the amplitude of the 1st companion indicates that small object is favored. The histogram of its period indicates that the period of the 1st companion is poorly constrained and there are many peaks in the histogram. The histogram of its eccentricity shows that the data almost provides no information for eccentricity of the 1st companion, and the posterior is very close to the prior for eccentricity given in Eqn.~(\ref{eq:prior-e}).}
 \label{fig:282-3-hist}
\end{figure}

\begin{figure}
 \centering
 \includegraphics[width=0.9\linewidth]{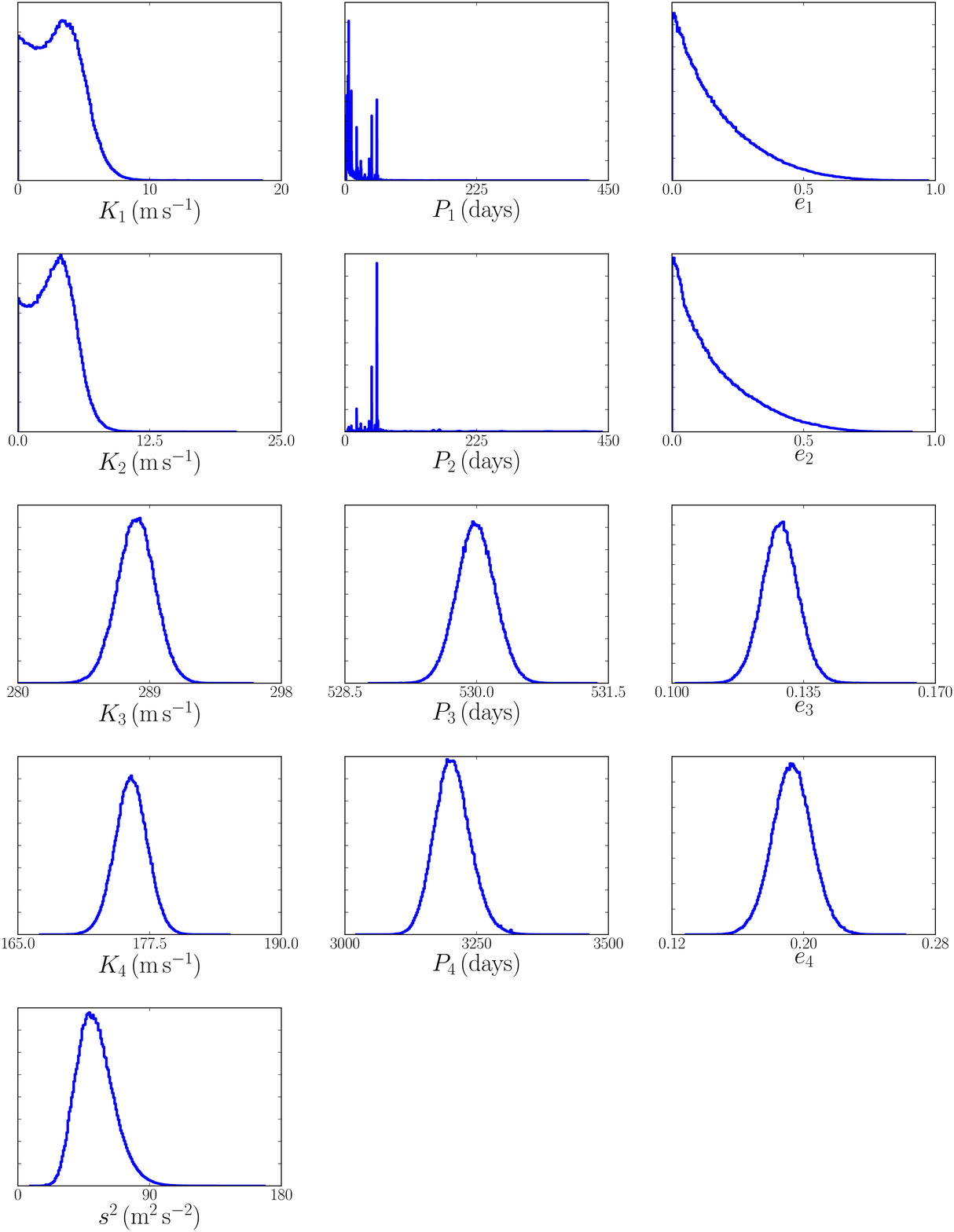}
 \caption{The posterior histograms of amplitudes, periods, eccentricities and jitter in the 4-companion model for RV data from HIP 88048 and $K_{\mathrm{min}} =0\,\unit{m\,s^{-1}}$. The histograms of the amplitudes of the 1st and 2nd companions indicate that small objects are favored. The histograms of their periods indicate that the periods are poorly constrained. The histograms of their eccentricities are both very similar to the prior for eccentricity given in Eqn.~(\ref{eq:prior-e}).}
 \label{fig:282-4-hist}
\end{figure}

We fit the RV data from HIP 88048 with 1, 2, 3 and 4-companion model. 
The fits and residuals from the four models are shown in Fig.~(\ref{fig:282-all-fits}) .
The 2-companion model gives a better fit than 1-companion.
But the 2 and 3-companion fits are nearly indistinguishable. 
The histograms of some of the parameters are shown in Fig.~(\ref{fig:282-2-hist}), Fig.~(\ref{fig:282-3-hist}), and Fig.~(\ref{fig:282-4-hist}). 
For 3 and 4-companion model, the histograms of the two large companions (large $K$) are very similar to the histograms from the 2-companion model. 
Note that the periods of the small companions (small $K$) in Fig.~(\ref{fig:282-3-hist}) and Fig.~(\ref{fig:282-4-hist}) are all badly constrained. 
But they do show some significant peaks. 
The histograms of small companions' eccentricities in Fig.~(\ref{fig:282-3-hist}) and Fig.~(\ref{fig:282-4-hist}) are very similar to the prior for eccentricity given in Eqn.~(\ref{eq:prior-e}).

The fully marginalized likelihoods of the 4 models given different $K_{\mathrm{min}}$ are shown in Tab.~(\ref{tab:282ZK}).
The logarithm of the FML of the 2, 3 and 4-companion models given different $K_{\mathrm{min}}$ are shown in Fig.~(\ref{fig:282lnZK}).
We set $C=0.01$ and use $(k+1)\cdot10^6$ samples to determine each $\Delta\beta$ for $k$-companion model. 
The error bars are obtained via Eqn.~(\ref{eq:z-var-est}).
Let $Z_{\mbox m\,k}$ represent the fully marginalized likelihood of $k$-companion model given some prior distribution.
We verified the error bar for $\widehat{Z}_{\mbox m\,1}$ given $K_{\mathrm{min}} = 0\, \unit{m\,s^{-1}}$ using $M=100$ independent evaluations.
This gave $\sigma_{\widehat{Z}_{\mbox m\,1}} =0.006\times10^{-356}$, which is consistent with 
the $0.005\times10^{-356}$ in Tab.~(\ref{tab:282ZK}).
All these results have the same unit $\left( \unit{m\,s^{-1}} \right)^{-N_{\mathrm{data}}}$, where $N_{\mathrm{data}}$ is the number of observations. The units are omitted to make the presentation clean. 

\subsubsection{Gliese 581}
\begin{figure}
 \centering
 \includegraphics[width=0.9\linewidth]{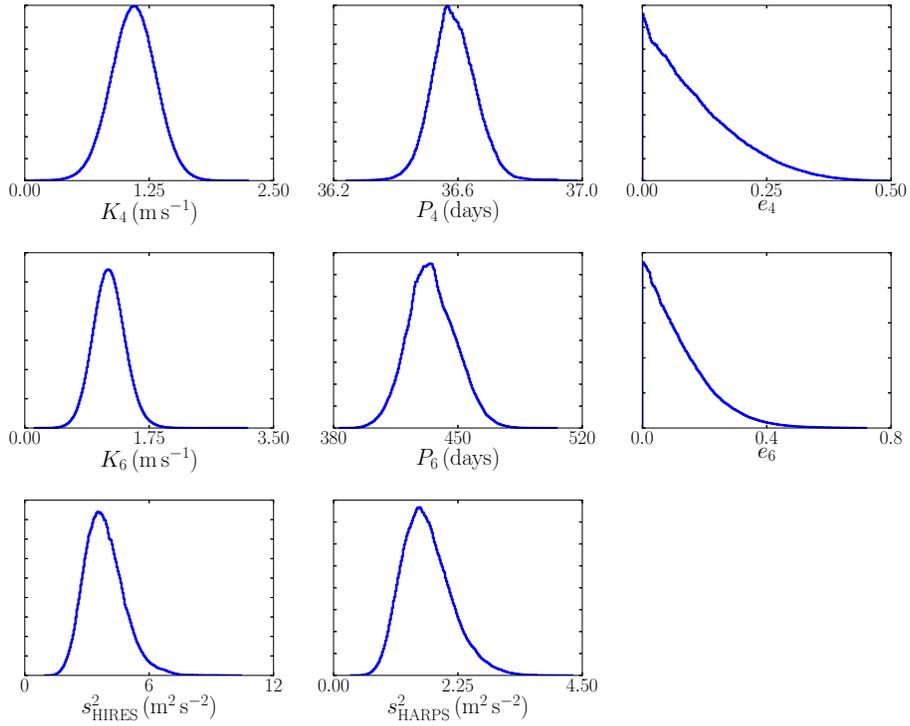}
 \caption{The posterior histograms of amplitudes, periods, eccentricities and jitters in the 6-planet model for RV data from Gliese 581. Only the histograms of the orbital parameters for the two planets which are not present in the 4-planet model are shown.}
 \label{fig:g581-6-2ls}
\end{figure}

We fit the combination of HARPS and HIRES data from Gliese 581 with 3, 4, 5 and 6-planet model. The histograms of the posterior of some of the parameters in the 6-planet model are shown in Fig.~(\ref{fig:g581-6-2ls}). The FML of these four models are
\begin{align}
Z_{\mbox m\,3} &= 3.5 \pm 0.1 \cdot 10^{-288}\;,\nonumber \\
Z_{\mbox m\,4} &= 2.9 \pm 0.1 \cdot 10^{-280}\;, \nonumber\\
Z_{\mbox m\,5} &= 2.5 \pm 0.2 \cdot 10^{-278}\;, \nonumber \\
Z_{\mbox m\,6} &= 2.1 \pm 0.2 \cdot 10^{-278}\; .
\label{eq:G581Z}
\end{align}
We set $C=0.01$ and use $(k+1)\cdot10^6$ samples to determine each $\Delta\beta$ for $k$-planet model, but we allow the number of samples to increase until we have a valid estimation of the auto-correlation time. Our estimated Bayes factor between between 5-planet and 6-planet model is far larger than reported in\cite{gregory11a}. We cannot confidently decide between 5 and 6-planet models solely based on Eqn.~(\ref{eq:G581Z}).

\subsection{Sensitivity to Priors}
\label{sec:sp}
For any Bayesian approach, the outcomes depend more or less on the choice of prior distributions. 
This is especially true in the case of the FML \citep{jaynes03a, cameron13a}. 
The results in Tab.~(\ref{tab:282ZK}) shows that the 2-companion model has the largest FML for all three $K_{\mathrm{min}}$.
But for different $K_{\mathrm{min}}$, the Bayes factors of the 2-companion model against other models are different. 
When $K_{\mathrm{min}}=0\, \unit{m\,s^{-1}}$, $Z_{\mbox m\,2}$ is only a little more than twice $Z_{\mbox m\,3}$.
But when $K_{\mathrm{min}}=5\, \unit{m\,s^{-1}}$, $Z_{\mbox m\,2}$ is more than 10 times $Z_{\mbox m\,3}$, 
and when $K_{\mathrm{min}}=10\, \unit{m\,s^{-1}}$, $Z_{\mbox m\,2}$ is more than 1000 times $Z_{\mbox m\,3}$. 
So we cannot rule out or confirm the existence of a 3rd companion in the system, but we can confidently rule out a 3rd companion which has a radial velocity contribution larger than $10\, \unit{m\,s^{-1}}$.

As for Gliese 581, Eqn.~(\ref{eq:G581Z}) shows that the Bayes factor between 5-planet and 6-planet model is about $1.2$.
We can then use this number and estimate the Bayes factor between these two models for priors with different $K_{\mathrm{max}}$.   
Since the maximum radial velocity from Gliese 581 in either HARPS or HIRES data set is less than $ 30\,\unit{m\,s^{-1}}$, 
the likelihood, Eqn.~(\ref{eq:mpl}), will be almost zero if the amplitude $K$ of any planet is larger than $30\,\unit{m\,s^{-1}}$.
So by changing $K_{\mathrm{max}}$ to any value larger than $30\,\unit{m\,s^{-1}}$, 
the only thing that will affect the FML is the normalization term in Eqn.~(\ref{eq:prior-K}).
For example, if we had chosen $K_{\mathrm{max}}=100\,\unit{m\,s^{-1}}$, the Bayes factor between 5-planet and 6-planet model would become $\log{(110/10)}/\log{(1010/10)}\cdot1.2\approx0.6$. On the other hand, if $K_{\mathrm{max}}=10000\,\unit{m\,s^{-1}}$, the Bayes factor between the two models would become $\log{(10010/10)}/\log{(1010/10)}\cdot1.2\approx1.8$. Which model is favored by the FML is very sensitive to the choice of prior.

\section{Discussion}
\label{sec:discussion}
In this paper, we implement geometric-path Monte Carlo to evaluate the fully marginalized likelihood of various models. We find that the estimator of the FML in geometric-path Monte Carlo is nearly unbiased, and the estimation of an uncertainty or error bar is straightforward and reasonable. We apply the algorithm to a Rosenbrock trial problem, our estimates of the fully marginalized likelihood match the value from  direct quadrature and the error bar match the results from repeated runs. We are also able to obtain the FML of multi-companion models for radial velocity data from HIP 88048 and Gliese 581 given various prior distributions.

The geometric-path Monte Carlo algorithm is a very fast method considering the challenging nature of the fully marginalized likelihood. For HIP 88048, the computation time for $Z_{\mbox m\,1}$ and $Z_{\mbox m\,2}$ is within hours. The computation of $Z_{\mbox m\,3}$ can be finished within a day, and $Z_{\mbox m\,4}$ a little more than a day. All the above computations were  done using a present-day workstation  machine with a single core. 

There are two important tuning parameters in geometric-path Monte Carlo, $C$ and $N$. Currently, our choice of them is somewhat arbitrary. It may be possible to find optimal $N$ and $C$ so that it takes the minimum computation time to achieve a desired uncertainty. But this is not trivial. For example, having a smaller $C$ and keeping $N$ unchanged does not guarantee a smaller uncertainty, because it may take more $\Delta\beta$ steps which increases the uncertainty and more $\Delta\beta$ steps also mean more burn-in. In practice, one may need to do a few trial runs to find a reasonable $C$ and $N$. 

The performance of the geometric-path Monte Carlo is constrained by the performance of the MCMC sampler one uses. 
When sampling difficult $p_\beta$, it may be very challenging to get a good estimation of the auto-correlation time $\tau$. 
In such cases, decreasing step size may improve the performance.
Currently we do not have a good understanding of the dependence of $\tau$ on $\Delta\beta$.
From what we have observed, $\tau$ of the chain of $Y^{\Delta\beta}$ is generally larger than that of the chain of $Y$,
and sometimes by a large amount.
Because $\Delta\beta$ is unknown and to be obtained from the procedure, we need to make sure that we have a valid estimation of $\tau$ of $Y^{\Delta\beta}$ for a wide range of $\Delta\beta$.

Our results for Gliese 581 are different from previous ones. 
This could be due to several reasons. 
First of all, we have different priors.
From  section~(\ref{sec:sp}), we can see that the FML is very sensitive to choice of priors. 
Second, the FML results in \cite{gregory11a} are evaluated using only HARPS data,
while we use the combination of both HARPS and HIRES data.
Third, there could be sampling issues with either of our evaluations that we are not aware of.
For our method, potential sampling problems usually reveal themselves when $\tau$ can not be confidently estimated.
But we have reasonable estimations of $\tau$ for all our evaluations.

\begin{figure}
 \centering
 \includegraphics[width=0.99\linewidth]{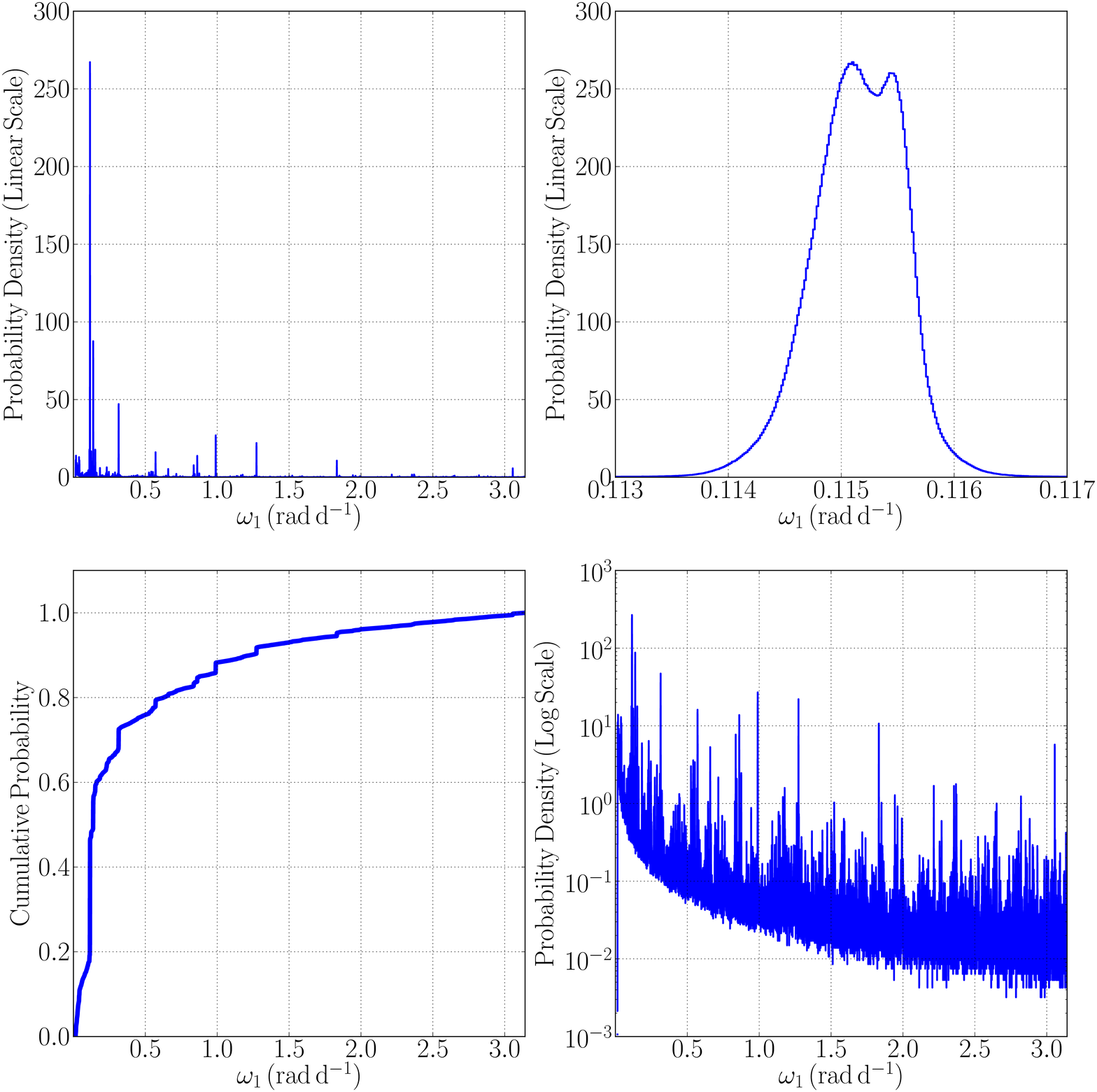}
 \caption{The figure on top left shows the fully resolved histogram of the posterior of $\omega_1$ in 3-companion model for RV data from HIP 88048 on linear scale. The figure on bottom right shows the same histogram on log scale. The figure on top right shows in detail the highest peak at around 54-day period. The cumulative probability figure on bottom left shows that the highest peak occupies considerable amount of probability.}
 \label{fig:282-3-hist-zi}
\end{figure}

\begin{figure}
 \centering
 \includegraphics[width=0.99\linewidth]{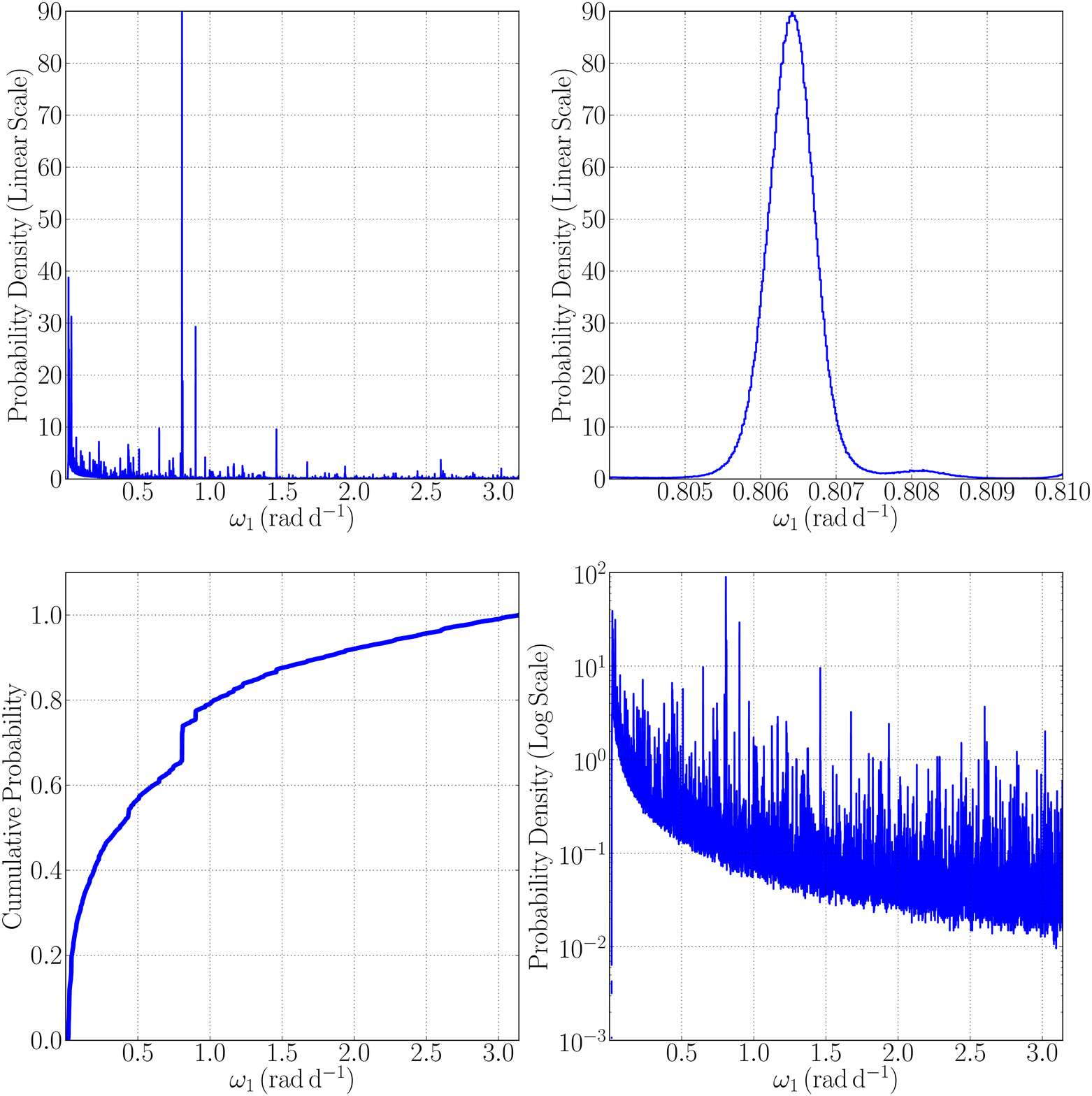}
 \caption{The figure on top left shows the fully resolved histogram of the posterior of $\omega_1$ in 3-companion model for fake RV data on linear scale. The figure on bottom right shows the same histogram on log scale. The figure on top right shows in detail the highest peak at around 8-day period. The cumulative probability figure on bottom left shows that the highest peak occupies very limited amount of probability. (The fake data is made with two companions, the parameters of which are taken from optimum 2-companion fit for RV data from HIP 88048, plus independent Gaussian noise, the standard deviation of which is the standard deviation of the real RV data residual of the optimum 2-companion fit.)}
 \label{fig:282-3-hist-zi-f}
\end{figure}

A detailed look at the computations for HIP 88048 makes it clear that there may be drawbacks to the 
Bayesian FML approach to estimating the number of planets or companions.
Fig.~(\ref{fig:282-3-hist-zi}) shows a histogram of the orbital angular speed of a possible third planet/companion.
This shows a very strong signal with an approximately 54-day period.
Fig.~(\ref{fig:282-3-hist-zi}) also shows that there is considerable probability in this narrow peak.
We have conducted computational experiments with fake data that contains signals only from the
two ``confirmed'' companions and additive Gaussian independent ``measurement'' error.
Fig.~(\ref{fig:282-3-hist-zi-f}) shows surprising narrow peaks in the posterior period for a non-existent
third companion, but that these peaks contain very little probability.
It is possible that a frequentist, hypothesis testing approach would find the 54-day period statistically
significant even though the Bayesian approach, with the priors we used, gives the 3-companion model 
less evidence than the 2-companion model.

\acknowledgements We thank Brendon Brewer (University of Auckland), Daniel Foreman-Mackey (NYU), Ross Fadely (NYU) for valuable discussions. We want to particularly thank Andreas Quirrenbach (University of Heidelberg) and Christian Schwab (Yale) for generously sharing data with us. We would also like to thank Ewan Cameron (Oxford University) for helpful comments. Partial support for FH and DWH was provided by NASA grant NNX12AI50G and NSF grant IIS-1124794. Partial support for JG was provided by the DOE office of Advanced Scientific Computing under grant DE-FG02-88ER25053.

\begin{table}
\centering
\begin{tabular}{lrr}
\hline
 & HIP 88048 b & HIP 88048 c \\
\hline
$K\,(\unit{m}\,\unit{s}^{-1})$ & $288.1\pm1.3$ & $175.8\pm 1.6$ \\
$P\,(\unit{day})$ & $529.9\pm0.2$ & $3211\pm 35$ \\
$\phi\,(\unit{rad})$ & $4.130\pm0.032$ & $3.859\pm0.046$ \\
$e$ & $ 0.1298 \pm 0.0045$ & $ 0.195 \pm 0.012$ \\
$\varpi\,(\unit{rad})$ & $1.732\pm0.032$ & $1.768\pm 0.039$ \\
$m\,\sin{i}\,(M_J)$ & $23.9 \pm 0.6$ & $26.3 \pm 0.7$\\
$a\,(\unit{AU})$ & $1.86 \pm 0.01$ & $6.17 \pm 0.04$\\
\hline
\end{tabular}
\begin{center}
\caption{Parameters for the two companions of HIP 88048 in 2-companion model fit.\label{tab:282param}}
\end{center}
\end{table}

\begin{table}
\centering
\begin{tabular}{lrrrr}
\hline
 $K_{\mbox{min}}$ & 1-Companion Model  & 2-Companion Model & 3-Companion Model & 4-Companion Model \\
\hline
$0\,\mathrm{m\,s^{-1}}$ &    $3.72 \pm 0.04 \cdot 10^{-356} $   &   $  1.98 \pm 0.02 \cdot 10^{-237}$    &   $7.4\pm0.1\cdot10^{-238}$   &   $1.47\pm0.03\cdot10^{-238}$ \\
$5\,\mathrm{m\,s^{-1}}$ &  $3.92 \pm 0.04 \cdot 10^{-356} $    &   $  2.23 \pm 0.02 \cdot 10^{-237}$   &  $1.27\pm0.01\cdot10^{-238}$   &  $8.6\pm0.2\cdot10^{-241}$ \\
$10\,\mathrm{m\,s^{-1}}$ &     $4.13 \pm 0.04\cdot 10^{-356} $    &   $  2.46 \pm 0.02 \cdot 10^{-237} $     &  $1.07\pm0.01\cdot10^{-240}$  &   $5.6\pm0.1\cdot10^{-248}$\\
\hline
\end{tabular}
\begin{center}
\caption{Marginalized Likelihood of 1, 2, 3 and 4-companion models for RV data from HIP 88048 with respect to different minimum velocity amplitude in prior.}\label{tab:282ZK}
\end{center}
\end{table}

\clearpage

\end{document}